\documentclass[aps,twocolumn,showpacs]{revtex4}
\usepackage{graphicx}
\usepackage{amsmath}
\usepackage{amssymb}

\begin{document}

\title{First-order integrability-breaking phase transitions in dynamical systems}

\author{Anne K\'etri P. da Fonseca$^{1,2}$, Marcelo de Almeida Presotto$^2$, Diego F. M. Oliveira$^1$, Edson D.\ Leonel$^2$}

\affiliation{$^1$School of Electrical Engineering and Computer Science, University of North Dakota, USA\\
$^2$Departamento de F\'isica, Unesp - Universidade Estadual Paulista - 
Av.24A. 1515, 13506-900, Rio Claro, SP, Brazil}

\date{\today} \widetext


\begin{abstract}
We investigate a discontinuous route from integrability to chaos using a confined stochastic random walk and a deterministic stadium-like billiard. In both systems, the stationary diffusive observable exhibits a finite jump at the transition: it vanishes at the unperturbed limit but approaches a finite, geometry-controlled value for arbitrarily small nonzero perturbations, providing the characteristic order-parameter signature of a first-order transition. Despite this discontinuity, the relaxation timescale diverges as the transition is approached, revealing critical slowing down. Both models exhibit normal diffusion with $\beta=1/2$, a perturbation-independent stationary state with $\alpha=0$, and a crossover iteration scaling as $n_x\propto\lambda^{-2}$, yielding $z=-2$, where $\lambda$ denotes the corresponding perturbation parameter. The common exponent set $(\alpha,\beta,z)=(0,1/2,-2)$ originates from the same coarse-grained mechanism: normal diffusion within a finite accessible domain with a diffusion coefficient that vanishes quadratically at the transition. The agreement between stochastic transport and deterministic chaotic scattering provides strong evidence for a common class of discontinuous dynamical transitions and extends the statistical-mechanics description of phase transitions to integrability-breaking dynamics.
\end{abstract}

\maketitle


\section{Introduction}
\label{introduction}

The concept of phase transitions originated in classical thermodynamics, with early developments such as Black's formulation of latent heat \cite{black} and Gibbs's theory of heterogeneous equilibrium \cite{gibbs}. A systematic classification was later introduced by Ehrenfest in terms of discontinuities in derivatives of thermodynamic potentials \cite{ehrenfest1933,ehrenfest2}. The subsequent observation of critical singularities beyond this thermodynamic description motivated the development of the modern statistical-mechanics framework of phase transitions \cite{moreissame}. A central element of this framework is Landau's concept of an order parameter, a macroscopic quantity whose behavior distinguishes different phases of a system \cite{landaunature}. In a first-order transition, the order parameter changes discontinuously at the transition point, whereas in a continuous (second-order) transition it evolves continuously across criticality. Modern descriptions further incorporate concepts such as symmetry breaking, topological defects, fluctuations, and critical scaling \cite{sethna2021statistical}. These ideas have proved remarkably general, providing a common language for phenomena ranging from ferromagnetism \cite{khanna1991magnetic,grinstein1976ferromagnetic} and superconductivity \cite{bianchi2002first,vojta2000quantum} to social \cite{social,social2} and biological systems \cite{bio1,bio2}, as well as nonlinear dynamical systems \cite{denisnovo,nontwist}.

In nonlinear dynamics, the transition from integrability to nonintegrability provides a particularly suitable setting in which to investigate critical phenomena. A broad class of systems, including one-dimensional maps near bifurcations \cite{bif}, the standard map \cite{leonel2020characterization}, and time-dependent collision models \cite{11nova,oliveira2013some}, exhibits scaling behavior when a control parameter drives the dynamics away from an integrable limit. For Hamiltonian systems, this scenario can be represented generically by
\begin{equation}
H(I_1,\theta_1,I_2,\theta_2)
=
H_0(I_1,I_2)
+
\epsilon H_1(I_1,\theta_1,I_2,\theta_2),
\end{equation}
where $H_0$ denotes the integrable contribution and $\epsilon$ controls the perturbation. At $\epsilon=0$, the dynamics is integrable and the phase space is organized by invariant structures associated with regular motion. For sufficiently small $\epsilon\neq0$, resonances generate chaotic layers embedded in a mixed phase space, where regular islands and chaotic regions coexist and transport remains constrained by invariant curves \cite{leonel2015dynamical}. As the perturbation increases, these barriers can be progressively destroyed \cite{MEISS}, eventually allowing global chaotic transport.

Close to the integrable limit, the diffusion inside the chaotic region commonly exhibits scaling invariance \cite{pathria2011statistical}. The dynamics is characterized by an initial growth regime, followed by a crossover at a characteristic iteration $n_x$ and, when the accessible chaotic region is bounded, by a stationary saturation regime. Scaling hypotheses expressed through a generalized homogeneous function connect these regimes and yield relations among the corresponding critical exponents. Within this framework, the stationary value of the diffusive observable provides a natural order parameter for the integrability-breaking transition. In the standard map, for example, this role is played by the saturation value of the root-mean-squared action $I_{\rm rms}$ \cite{kenji}, whereas in time-dependent collision models it is associated with the root-mean-squared particle velocity $V_{\rm rms}$ \cite{oliveira2013some,daFonseca2025PRE}. In these systems, the stationary observable approaches zero continuously as $\epsilon\rightarrow0$, while the associated susceptibility diverges, establishing a continuous, or second-order, integrability-breaking transition.

This raises a fundamental question: \emph{must the transition from integrability to nonintegrability always be continuous?} Here we show that the answer is no. We investigate two systems with very different microscopic dynamics: (i) a stochastic random walk confined to a finite interval and (ii) a deterministic two-dimensional stadium-like billiard whose boundary deformation controls the transition between integrable and chaotic dynamics. Despite their different microscopic origins, both systems display the same striking scenario. At the critical point, the stationary diffusive observable vanishes, whereas for an arbitrarily small but finite value of the control parameter it approaches a nonzero value that is independent of the distance from criticality. The corresponding order parameter therefore exhibits a finite jump at the transition, providing a dynamical realization of a first-order integrability-breaking phase transition.

Remarkably, this discontinuity of the asymptotic order parameter coexists with a divergent dynamical timescale. As the control parameter approaches its critical value, the crossover iteration $n_x$, which determines the number of iterations required for the dynamics to reach the stationary regime, diverges according to a power law. The system therefore exhibits critical slowing down even though its stationary order parameter changes discontinuously. Slow dynamical responses and diverging transient timescales are known to arise near transitions, bifurcations, and catastrophic shifts in nonlinear and complex systems \cite{vannes2007slow,scheffer2009early,kuehn2011mathematical}. Here, however, the combination of a discontinuous stationary order parameter with a scaling divergence of $n_x$ provides the central signature of the transition. We further show that the random-walk and billiard models share the same scaling structure and critical exponents, suggesting that they belong to the same dynamical universality class.

The remainder of this paper is organized as follows. Section \ref{xsec2} introduces the one-dimensional random-walk model and develops its scaling description in terms of the root-mean-squared displacement $x_{\mathrm{rms}}$. Section \ref{xsec3} extends the analysis to the two-dimensional stadium-like billiard, where the corresponding dispersion is quantified by the roughness $\omega$. In Sec. \ref{xsec4}, we characterize the transition within a statistical-mechanics framework, identify the order parameter, and compare the scaling properties and critical exponents of the two systems. Finally, Sec. \ref{xsec5} summarizes our main results and conclusions.

\section{One-dimensional model: random walk in a line}
\label{xsec2}

We begin with a minimal stochastic model that isolates the diffusive mechanism underlying the transition investigated in this work. Consider an ensemble of independent random walkers confined to a finite interval $-L\leq x\leq L$, with perfectly reflecting boundaries at $x=\pm L$. At each iteration, the position of a walker evolves according to
\begin{equation}
    x_{n+1}=x_n+\varepsilon Z(n),
    \label{Eq1}
\end{equation}
where $\varepsilon$ controls the maximum step amplitude and $Z(n)$ is an independent random variable uniformly distributed in the interval $[-1,1]$. The standard one-dimensional random walk provides one of the simplest descriptions of diffusive transport in statistical physics \cite{randomwalk}. Here, however, the finite boundaries introduce a stationary regime that allows us to define an asymptotic macroscopic observable and investigate its behavior as $\varepsilon\rightarrow0$.

\begin{figure}[h]
  \centering
    \centerline{\includegraphics[width=\linewidth]{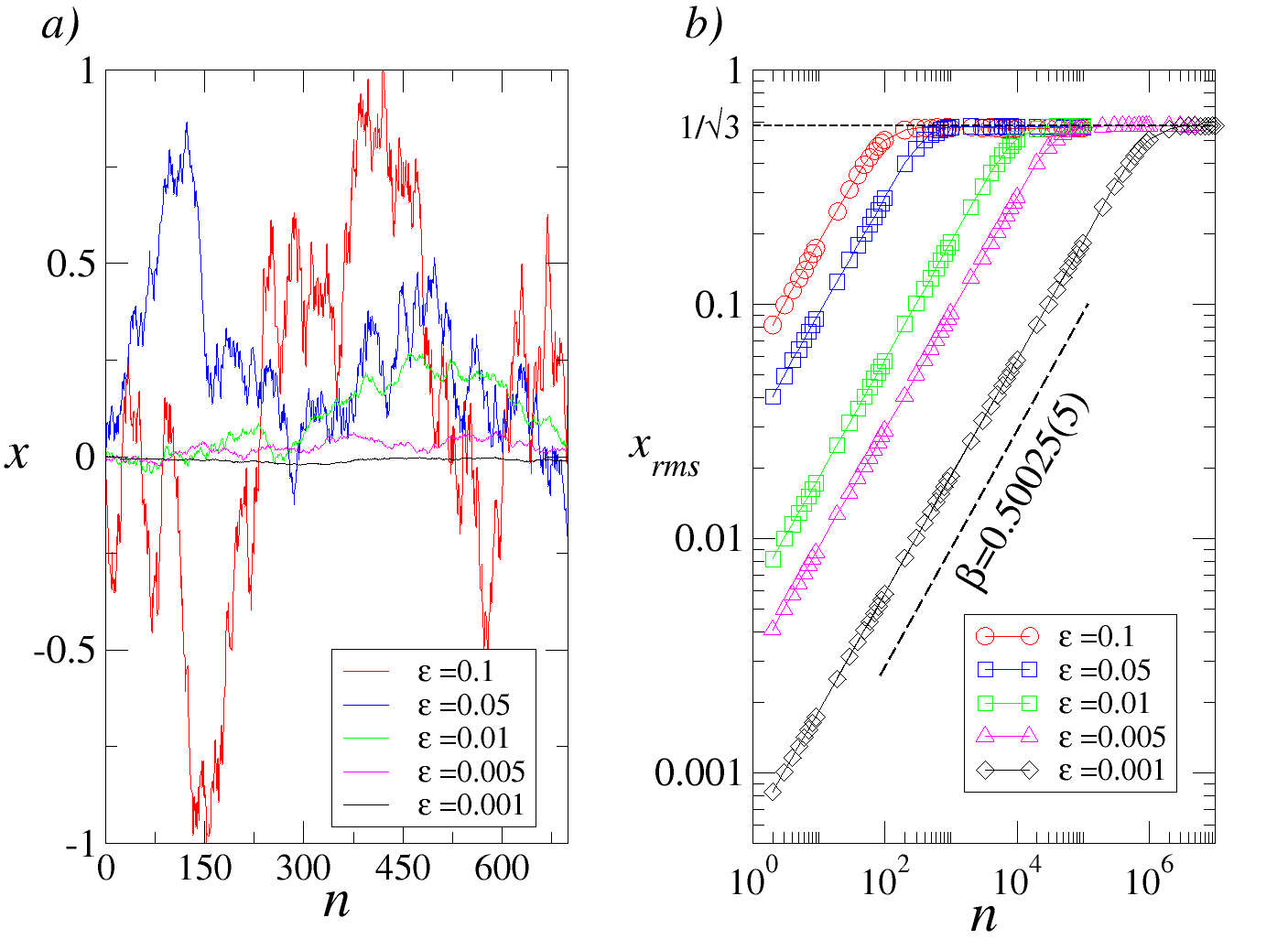}}
  \caption{Five representative trajectories $x$ vs. $n$ (a) and the root-mean-squared displacement $x_{\mathrm{rms}}$ vs. $n$ (b), obtained from Eq.~(\ref{Eq1}) for an ensemble of $5000$ particles and different values of $\varepsilon \in [10^{-3},10^{-1}]$.}
  \label{Fig1}
\end{figure}

Figure~\ref{Fig1}(a) shows representative trajectories obtained from Eq.~(\ref{Eq1}) for different values of $\varepsilon$. For any $\varepsilon\neq0$, the walkers progressively explore the available interval, whereas decreasing $\varepsilon$ substantially increases the number of iterations required to reach the boundaries and sample the entire accessible region.

To quantify this evolution, we introduce the time- and ensemble-averaged root-mean-squared displacement
\begin{equation}
    x_{\mathrm{rms}}(n)=
    \sqrt{\frac{1}{M}\sum^{M}_{i=1}
    \frac{1}{n}\sum^{n}_{j=1}x^2_{i,j}},
    \label{Eq2}
\end{equation}
where $M$ is the number of trajectories and $x_{i,j}$ denotes the position of the $i$th walker at iteration $j$. Figure~\ref{Fig1}(b) shows $x_{\mathrm{rms}}(n)$ for an ensemble of $5000$ particles and different values of $\varepsilon$.

Two dynamical regimes are clearly identified. For $n\ll n_x$, the walkers have not yet explored the entire interval and
\begin{equation}
x_{\mathrm{rms}}\propto n^\beta,
\end{equation}
with $\beta\simeq1/2$, as expected for normal diffusion. For $n\gg n_x$, the reflecting boundaries confine the dynamics and the probability distribution approaches the uniform stationary distribution in $[-L,L]$. Consequently,
\begin{equation}
x_{\mathrm{sat}}
=
\left[
\frac{1}{2L}\int_{-L}^{L}x^2\,dx
\right]^{1/2}
=
\frac{L}{\sqrt{3}},
\end{equation}
independently of $\varepsilon$ for every $\varepsilon\neq0$. The scaling
$x_{\mathrm{sat}}\propto\varepsilon^\alpha$ therefore yields
$\alpha=0$.

\begin{figure}[h!]
  \centering
    \centerline{\includegraphics[width=0.95\linewidth]{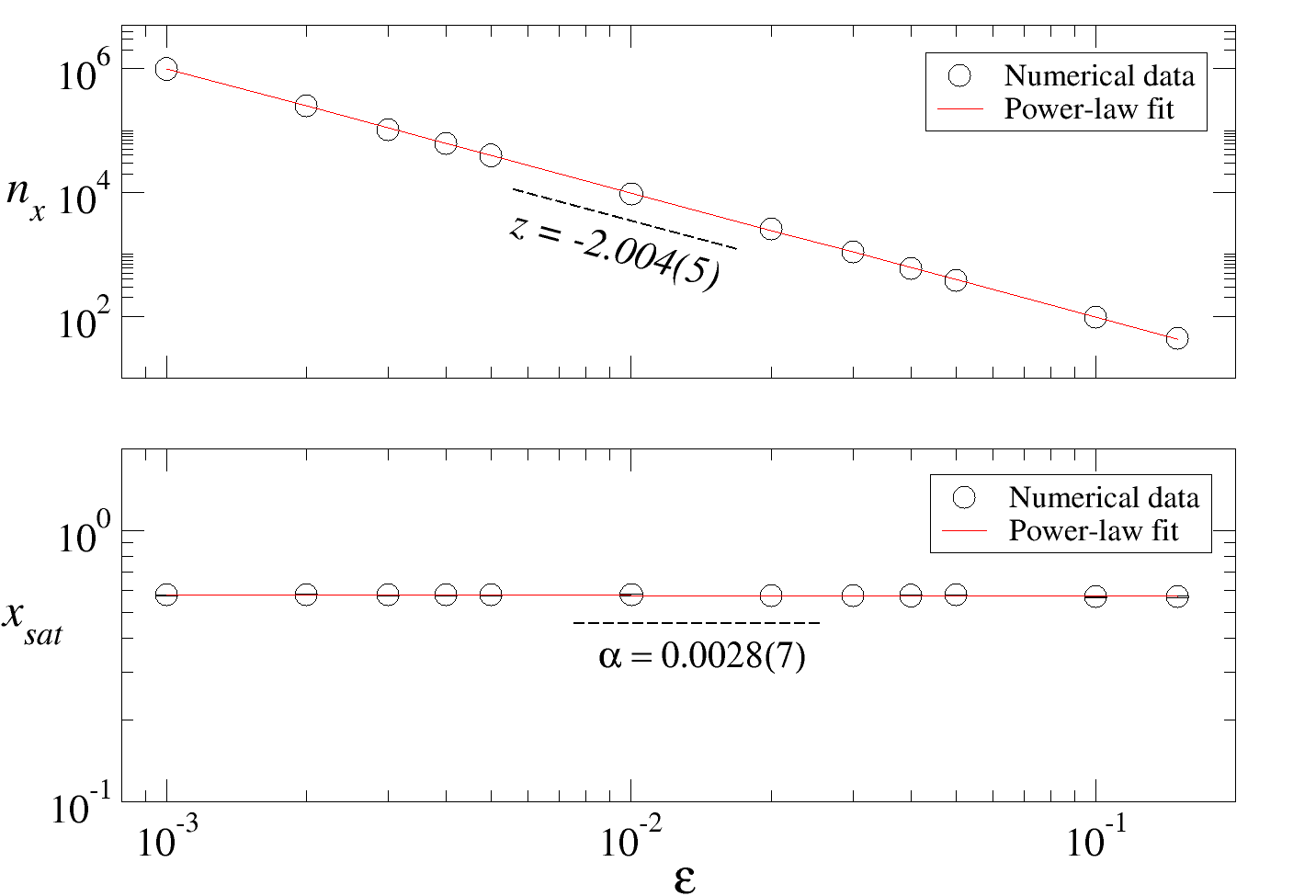}}
  \caption{Dependence of the crossover iteration $n_x$ and stationary displacement $x_{\mathrm{sat}}$ on the control parameter $\varepsilon$. Power-law fits yield $z=-2.00(5)$ and $\alpha=0.00(2)$, respectively.}
  \label{Fig2}
\end{figure}

The crossover iteration separating the diffusive growth and stationary regimes scales as
\begin{equation}
n_x\propto\varepsilon^z.
\end{equation}
Figure~\ref{Fig2} shows $x_{\mathrm{sat}}$ and $n_x$ as functions of $\varepsilon$. Power-law fits yield $z=-2.00(5)$ and $\alpha=0.00(2)$, in agreement with the expected values $z=-2$ and $\alpha=0$.

\begin{figure}[h!]
  \centering
    \centerline{\includegraphics[width=0.95\linewidth]{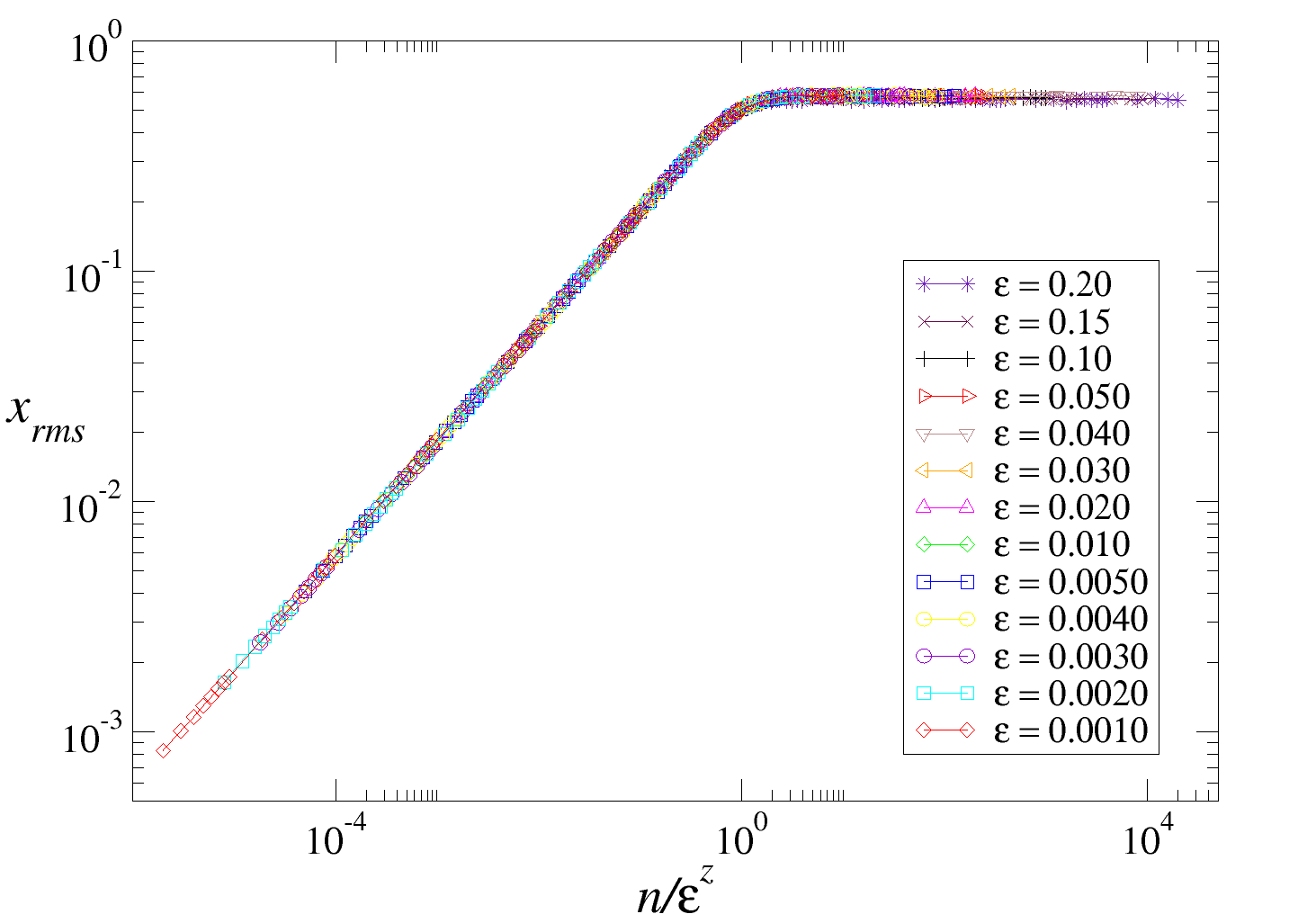}}
  \caption{Scaling collapse of $x_{\mathrm{rms}}$ as a function of $n/\varepsilon^z$ for $z=-2$, an ensemble of $5000$ particles, and $\varepsilon \in [1\times10^{-3},2\times10^{-1}]$.}
  \label{Fig3}
\end{figure}

The scaling behavior can be further tested by rescaling the iteration number according to $n\rightarrow n/\varepsilon^z$. As shown in Fig.~\ref{Fig3}, the curves obtained for different values of $\varepsilon$ collapse onto a single scaling function when $z=-2$.

The singular character of the limit $\varepsilon\rightarrow0$ is particularly important. For every finite $\varepsilon$, no matter how small, the long-time dynamics explores the entire accessible interval and therefore
\begin{equation}
\lim_{n\rightarrow\infty}x_{\mathrm{rms}}(n,\varepsilon)
=
\frac{L}{\sqrt{3}},
\qquad \varepsilon\neq0.
\end{equation}
At $\varepsilon=0$, however, the walkers remain permanently at their initial position and
\begin{equation}
x_{\mathrm{rms}}(n,0)=0.
\end{equation}
Thus,
\begin{equation}
x_{\mathrm{sat}}(\varepsilon)=
\begin{cases}
0, & \varepsilon=0,\\[2mm]
L/\sqrt{3}, & \varepsilon\neq0,
\end{cases}
\label{EqOrderRW}
\end{equation}
revealing a finite discontinuity of the stationary observable at the transition. At the same time, the number of iterations required to reach this stationary state diverges as $\varepsilon\rightarrow0$. This coexistence of a discontinuous asymptotic observable and a diverging dynamical timescale will be central to our characterization of the first-order transition in Sec.~\ref{xsec4}.

The stochastic nature of the model also permits an analytical description in the continuous limit. For $\varepsilon\ll L$, the evolution of the probability density $P(x,n)$ is governed by the diffusion equation
\begin{equation}
    \frac{\partial P(x,n)}{\partial n}
    =
    D\frac{\partial^2P(x,n)}{\partial x^2},
    \label{Eq3}
\end{equation}
where, for a uniformly distributed $Z(n)\in[-1,1]$,
\begin{equation}
D=\frac{\langle(\Delta x)^2\rangle}{2}
=\frac{\varepsilon^2\langle Z^2\rangle}{2}
=\frac{\varepsilon^2}{6}.
\label{EqDiffCoeff}
\end{equation}
The reflecting boundaries impose the Neumann conditions
\begin{equation}
\left.
\frac{\partial P(x,n)}{\partial x}
\right|_{x=\pm L}=0,
\end{equation}
while the initial condition $P(x,0)=\delta(x)$ places all walkers at the center of the interval.

Using separation of variables, the solution is
\begin{equation}
    P(x,n) = \frac{1}{2L}
    + \frac{1}{L}\sum_{k=1}^{\infty}
    \cos\left(\frac{k\pi x}{L}\right)
    \exp\left(
    -\frac{k^2\pi^2 D n}{L^2}
    \right).
    \label{Eq4}
\end{equation}
The uniform term $1/(2L)$ represents the stationary distribution, whereas the remaining modes decay exponentially with characteristic times proportional to $L^2/D$. In particular, the slowest mode, $k=1$, defines the relaxation timescale
\begin{equation}
\tau_1=\frac{L^2}{\pi^2D}
\propto\varepsilon^{-2},
\label{EqRelaxRW}
\end{equation}
providing an analytical origin for the divergence of $n_x$.

\begin{figure}[h!]
  \centering
    \centerline{\includegraphics[width=0.95\linewidth]{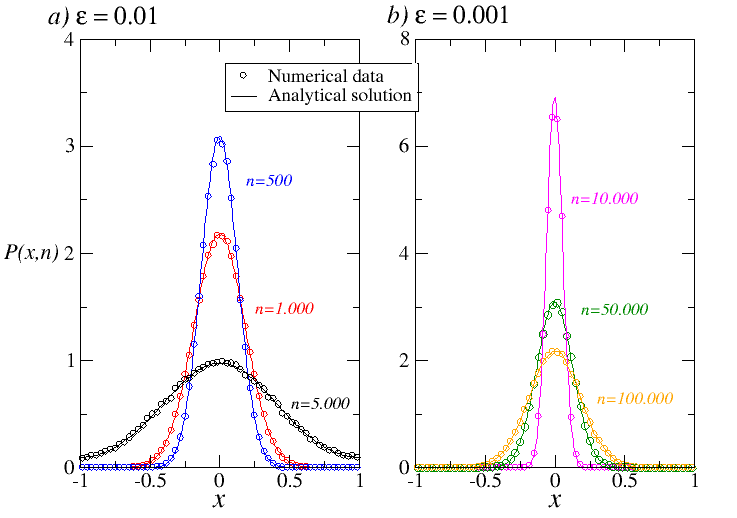}}
  \caption{Probability density $P(x)$ obtained numerically for an ensemble of $M=100\,000$ particles for (a) $\varepsilon=0.01$ and (b) $\varepsilon=0.001$ at the indicated iterations. Continuous lines show the analytical prediction of Eq.~(\ref{Eq4}).}
  \label{Fig4}
\end{figure}

Figure~\ref{Fig4} compares Eq.~(\ref{Eq4}) with numerical probability distributions obtained from an ensemble of $100\,000$ walkers. The agreement confirms that the continuous diffusion description accurately captures the dynamics in the regime considered.

The analytical solution also yields the mean-squared displacement at iteration $j$,
\begin{equation}
    \langle x^2 \rangle_j
    =
    \frac{L^2}{3}
    +
    \frac{4L^2}{\pi^2}
    \sum_{k=1}^{\infty}
    \frac{(-1)^k}{k^2}
    \exp\left(
    -\frac{k^2\pi^2Dj}{L^2}
    \right).
\end{equation}
Averaging this expression over the first $n$ iterations according to Eq.~(\ref{Eq2}) gives
\begin{equation}
\begin{aligned}
    x_{\mathrm{rms}}(n) = L & \left\{ \frac{1}{3}
    + \frac{4}{\pi^2}
    \sum_{k=1}^{\infty}\frac{(-1)^k}{k^2}
    \frac{1}{n}
    \exp\left(-\frac{k^2\pi^2D}{L^2}\right)
    \right.\\
    &\times\left.
    \left[
    \frac{
    1-\exp\left(-\frac{k^2\pi^2Dn}{L^2}\right)
    }{
    1-\exp\left(-\frac{k^2\pi^2D}{L^2}\right)
    }
    \right]
    \right\}^{1/2}.
\end{aligned}
\label{Eq5}
\end{equation}

Equation~(\ref{Eq5}) provides an analytical description of both limiting regimes. For $n\rightarrow\infty$, the transient contribution vanishes as $\mathcal{O}(1/n)$ and
\begin{equation}
x_{\mathrm{rms}}\rightarrow\frac{L}{\sqrt{3}},
\end{equation}
confirming $\alpha=0$. Before the boundaries become relevant, normal diffusion gives
\begin{equation}
\langle x^2(j)\rangle\simeq2Dj.
\end{equation}
Because Eq.~(\ref{Eq2}) includes a time average,
\begin{equation}
x_{\mathrm{rms}}^2(n)
\simeq
\frac{1}{n}\sum_{j=1}^{n}2Dj
\simeq Dn,
\end{equation}
and hence
\begin{equation}
x_{\mathrm{rms}}(n)\simeq\sqrt{Dn}
\propto\varepsilon n^{1/2}.
\end{equation}
Therefore $\beta=1/2$. Matching the diffusive and saturation regimes at $n=n_x$ gives
\begin{equation}
\varepsilon n_x^\beta\sim\frac{L}{\sqrt{3}},
\end{equation}
which immediately yields
\begin{equation}
n_x\propto\varepsilon^{-1/\beta}
=\varepsilon^{-2},
\end{equation}
and consequently
\begin{equation}
z=-\frac{1}{\beta}=-2.
\end{equation}

The same exponent relation follows from the scaling hypothesis
\begin{equation}
x_{\mathrm{rms}}(n,\varepsilon)
=
\varepsilon^\alpha
f\left(\frac{n}{\varepsilon^z}\right).
\end{equation}
For $u=n/\varepsilon^z\ll1$, assuming $f(u)\sim u^\beta$ gives
\begin{equation}
x_{\mathrm{rms}}
\sim
\varepsilon^{\alpha-z\beta}n^\beta.
\end{equation}
Comparison with the diffusive behavior
$x_{\mathrm{rms}}\propto\varepsilon n^\beta$
leads to
\begin{equation}
\alpha-z\beta=1.
\label{EqScalingRelationRW}
\end{equation}
Since $\alpha=0$ and $\beta=1/2$, Eq.~(\ref{EqScalingRelationRW}) yields $z=-2$, consistently with the numerical measurements and the analytical relaxation time of Eq.~(\ref{EqRelaxRW}).

The confined random walk therefore establishes a minimal stochastic reference for the transition studied here. Its stationary observable changes discontinuously at $\varepsilon=0$, while its relaxation timescale diverges continuously as $\varepsilon^{-2}$. In the following section, we show that the same scaling structure emerges in a deterministic two-dimensional chaotic billiard, despite the fundamentally different microscopic origin of its dynamics.

\section{Two-dimensional model: dispersing stadium billiard}
\label{xsec3}

We now investigate whether the scaling scenario identified in the stochastic model can emerge from a purely deterministic chaotic dynamics. Dynamical billiards provide a particularly suitable setting for this purpose because their global properties are controlled by the geometry of the boundary \cite{chernov2006chaotic}. A particle moves freely between successive collisions and undergoes specular reflections at the boundary, so that geometric deformations alone can drive the system from regular to chaotic dynamics.

\begin{figure}[h!]
  \centering
    \centerline{\includegraphics[width=\linewidth]{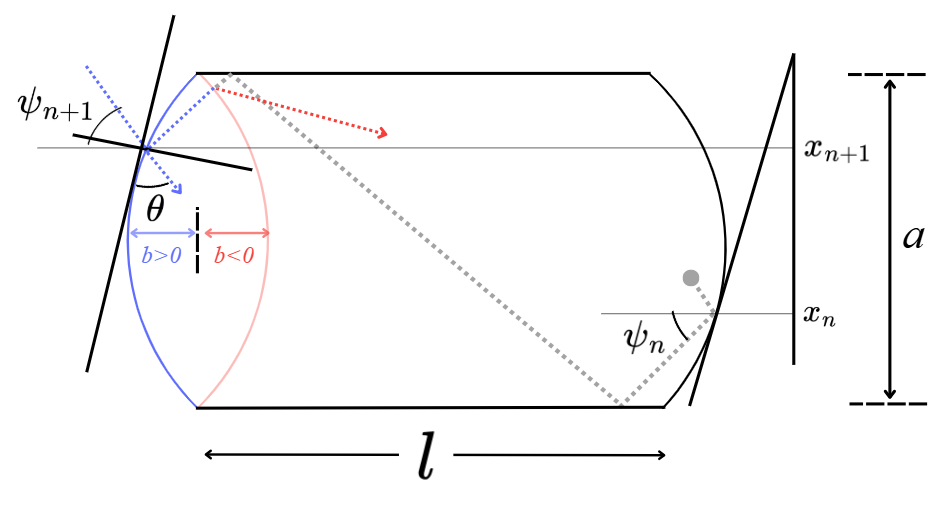}}
  \caption{Geometry of the stadium-like billiard and its relevant parameters. The focusing and dispersing configurations are distinguished by the sign of the geometric control parameter $b$.}
  \label{Fig5}
\end{figure}

The effect of the boundary curvature is particularly important. Dispersing components, as in the Lorentz gas and Sinai billiard, promote dynamical instability and exponential separation of nearby trajectories \cite{Sinai70}. Focusing boundaries, on the other hand, can support mixed phase spaces in which regular islands coexist with chaotic regions, as observed in oval billiards \cite{lopac2002chaotic} and stadium-like geometries \cite{Bunimovich1979}. Here we consider a generalized stadium billiard in which the semicircular components of the Bunimovich stadium are replaced by parabolic boundaries \cite{loskutov2002}, as illustrated in Fig.~\ref{Fig5}. The geometric parameter $b$ controls the curvature of the parabolic components and therefore plays the role of the perturbation parameter. For $b=0$, the curved components reduce to straight segments and the system becomes a rectangular, fully integrable billiard. According to the convention adopted here, $b>0$ corresponds to focusing boundaries, whereas $b<0$ defines the dispersing regime.

The two sides of the integrable limit exhibit qualitatively different routes to chaos. In the focusing regime ($b>0$), regular structures persist for finite values of the deformation and are progressively modified as $b$ increases \cite{loskutov2002,livorati2011family,daFonseca2026}. This produces a mixed phase space and a continuous variation of the corresponding stationary diffusive observable. In contrast, in the dispersing regime ($b<0$), arbitrarily small boundary curvature destabilizes the regular dynamics of the rectangular limit and produces a globally chaotic phase-space structure. This qualitative difference between the two sides of $b=0$ provides a natural setting for comparing continuous and discontinuous routes from integrability to chaos.

The dynamics can be described by the mapping
$T(\xi_n,\psi_n)=(\xi_{n+1},\psi_{n+1})$, where
$\xi_n=x_n/a\in[0,1)$ denotes the dimensionless collision coordinate and $\psi_n$ is the angle of the trajectory with respect to the vertical direction, measured clockwise. Using the unfolding construction to eliminate dynamically equivalent collisions \cite{chernov2006chaotic}, the parabolic boundaries can be written as
$f(x)=Ax^2+Bx+C$. Imposing $f(0)=f(a)=0$ and using the geometric parameters shown in Fig.~\ref{Fig5} gives
$A=\pm4b/a^2$, $B=\mp4b/a$, and $C=0$. The resulting map for the stadium-like billiard with static parabolic boundaries is \cite{livorati2011family}
\begin{equation}
    T: \begin{cases}
    \xi_{n+1} = (\xi_n + \frac{l}{a}\tan \psi_n) \mod 1, \\
    \psi_{n+1} = \psi_n \mp \frac{8|b|}{a}(2\xi_{n+1}-1),
    \end{cases}
\label{mapa}
\end{equation}
where the minus sign corresponds to the focusing geometry and the plus sign to the dispersing case.

\begin{figure*}[t!]
  \centering
    \centerline{\includegraphics[width=1.0\textwidth]{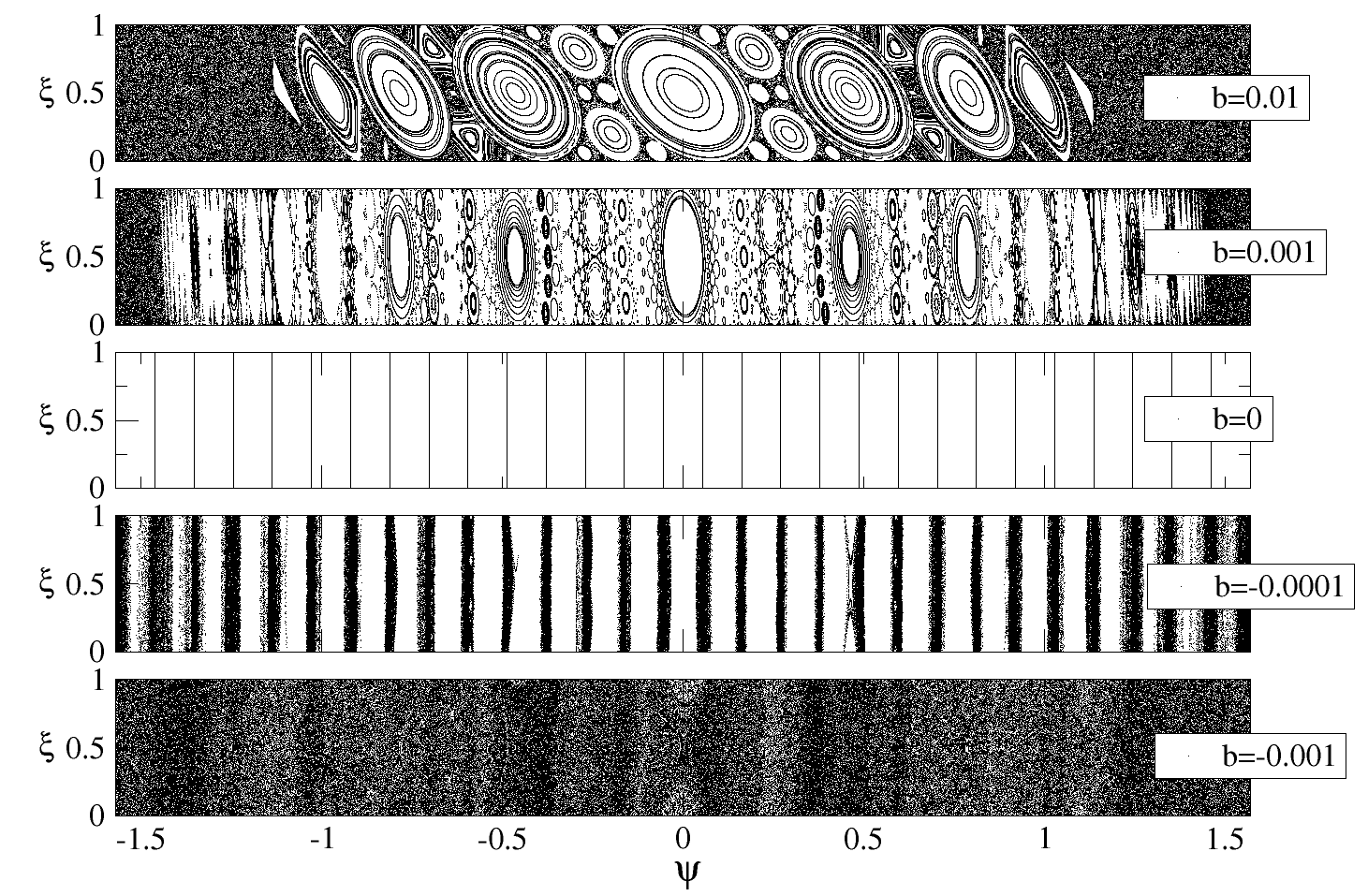}}
  \caption{Phase-space portraits of the stadium-like billiard for $a=0.5$, $l=1$, and the indicated values of the geometric parameter $b$, spanning the focusing and dispersing regimes.}
  \label{Fig6}
\end{figure*}

Figure~\ref{Fig6} shows representative phase-space portraits obtained for $a=0.5$, $l=1$, and different values of $b$. Each panel was constructed using an ensemble of $100$ initial conditions evolved for $n=10^3$ collisions. The focusing side exhibits the characteristic coexistence of regular and chaotic structures, whereas the dispersing side displays the abrupt emergence of a chaotic phase space from the integrable rectangular limit at $b=0$.

The distinction between the two regimes is also reflected in the stability of the periodic orbits. For the focusing geometry, a linear stability analysis yields the critical value
$b_c=a^2/(4l)$ associated with the destruction of the relevant regular structure \cite{daFonseca2026}. In the dispersing case, the trace of the Jacobian evaluated at the corresponding fixed points is
\begin{equation}
\operatorname{Tr}J
=
2+\frac{16|b|l}{a^2\cos^2\psi^*}.
\end{equation}
For any finite $b<0$, this expression satisfies $\operatorname{Tr}J>2$, so that the fixed points are hyperbolic. The dispersing perturbation therefore immediately introduces local dynamical instability as the system departs from the integrable limit. As we show below, the first-order character of the transition is established independently by the discontinuous behavior of the stationary macroscopic observable.

The transport generated by Eq.~(\ref{mapa}) occurs primarily in the angular variable $\psi$. To quantify its spreading, for each trajectory $j$ we define
\begin{equation}
\overline{\psi}_j(n,b)
=
\frac{1}{n}\sum_{i=1}^{n}\psi_{i,j},
\end{equation}
and the corresponding roughness
\begin{equation}
    \omega(n,b) =
    \frac{1}{M}\sum^{M}_{j=1}
    \sqrt{\overline{\psi^2}_j(n,b)-
    \overline{\psi}_j^2(n,b)}.
    \label{omega}
\end{equation}
This observable measures the average angular dispersion accumulated along the trajectories and provides the billiard counterpart of the root-mean-squared displacement introduced for the stochastic model.

\begin{figure}[h]
 \centering
   \centerline{\includegraphics[width=1.1\linewidth]{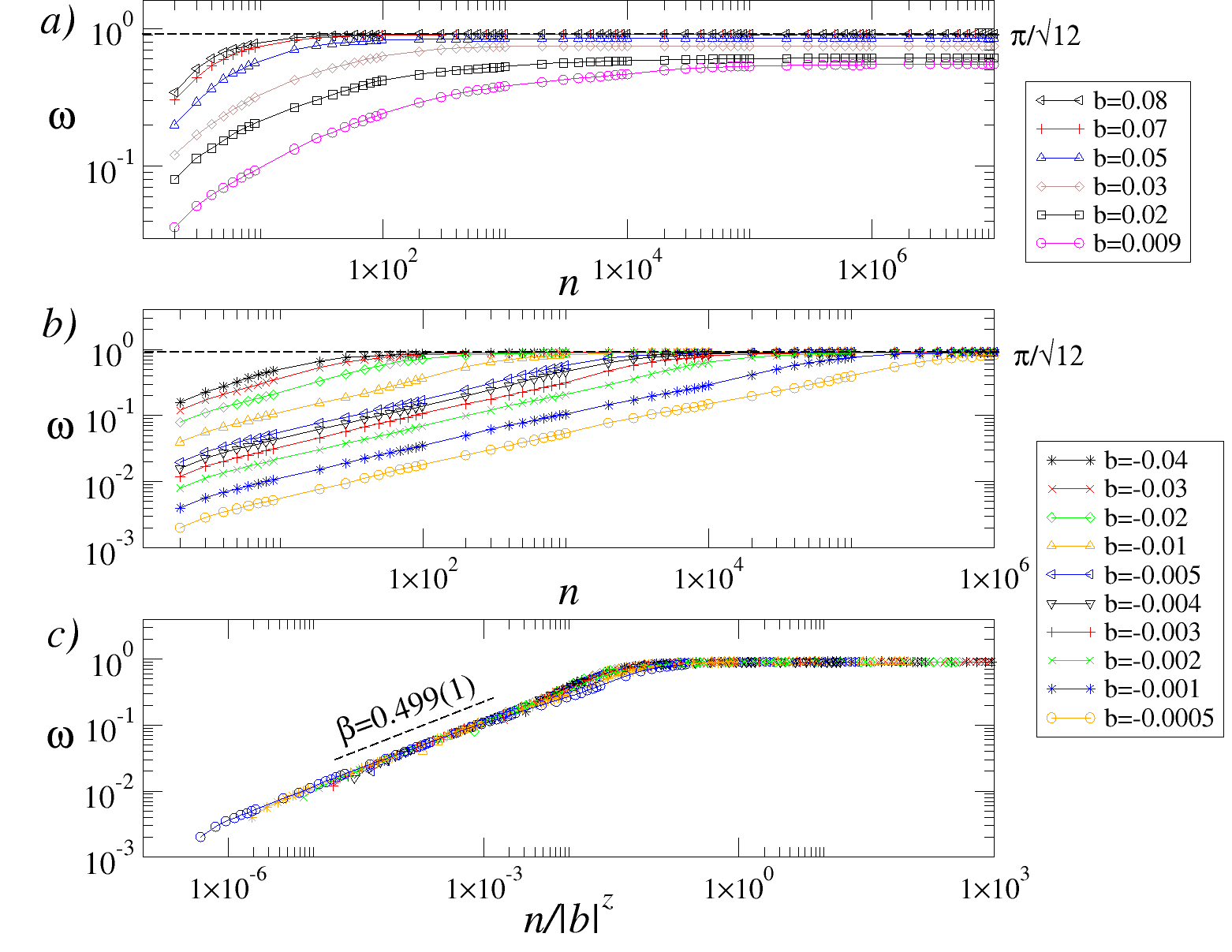}}
 \caption{Roughness $\omega(n)$ as a function of the number of collisions $n$ for different values of $b$. Panel (a) shows the focusing regime ($b>0$), while panel (b) corresponds to the dispersing regime ($b<0$). Panel (c) shows the scaling collapse of the dispersing data under the transformation $n\rightarrow n/|b|^z$ with $z=-2$.}
 \label{Fig7}
\end{figure}

Figure~\ref{Fig7} shows $\omega(n,b)$ for different values of the boundary deformation. The averages were computed over an ensemble of $10^3$ initial conditions uniformly distributed along $\xi_0\in[0,1]$, with the trajectories initialized in the chaotic component of phase space.

The focusing regime, shown in Fig.~\ref{Fig7}(a), provides a useful reference for the continuous transition previously investigated in this system \cite{daFonseca2026}. At short times, $\omega$ grows algebraically before crossing over to a stationary value $\omega_{\mathrm{sat}}$ that depends continuously on $b$. As the chaotic component expands, $\omega_{\mathrm{sat}}$ approaches the value
\begin{equation}
\omega_{\mathrm{erg}}
=
\frac{\pi}{\sqrt{12}}
\simeq0.9069,
\end{equation}
corresponding to the standard deviation of a uniform angular distribution over
$\psi\in[-\pi/2,\pi/2]$. Previous results showed that
$\omega_{\mathrm{sat}}\propto b^{\alpha'}$, with
$\alpha'\simeq0.28$, while the crossover iteration diverges with
$z\simeq-1.5$ \cite{daFonseca2026}. In particular,
$\omega_{\mathrm{sat}}\rightarrow0$ continuously as $b\rightarrow0^+$, providing the characteristic order-parameter behavior of a continuous transition.

A qualitatively different scenario emerges on the dispersing side. As shown in Fig.~\ref{Fig7}(b), for $n<n_x$ the angular roughness grows according to
\begin{equation}
\omega(n,b)\propto n^\beta,
\end{equation}
with $\beta\simeq1/2$, indicating normal angular diffusion. For
$n>n_x$, all curves converge to the same stationary value
\begin{equation}
\omega_{\mathrm{sat}}
\simeq
\frac{\pi}{\sqrt{12}},
\end{equation}
independently of $|b|$. The stationary angular distribution is therefore consistent with the uniform distribution over the accessible angular interval. Consequently,
\begin{equation}
\omega_{\mathrm{sat}}\propto |b|^\alpha
\end{equation}
yields $\alpha=0$.

\begin{figure}[h!]
  \centering
    \centerline{\includegraphics[width=\linewidth]{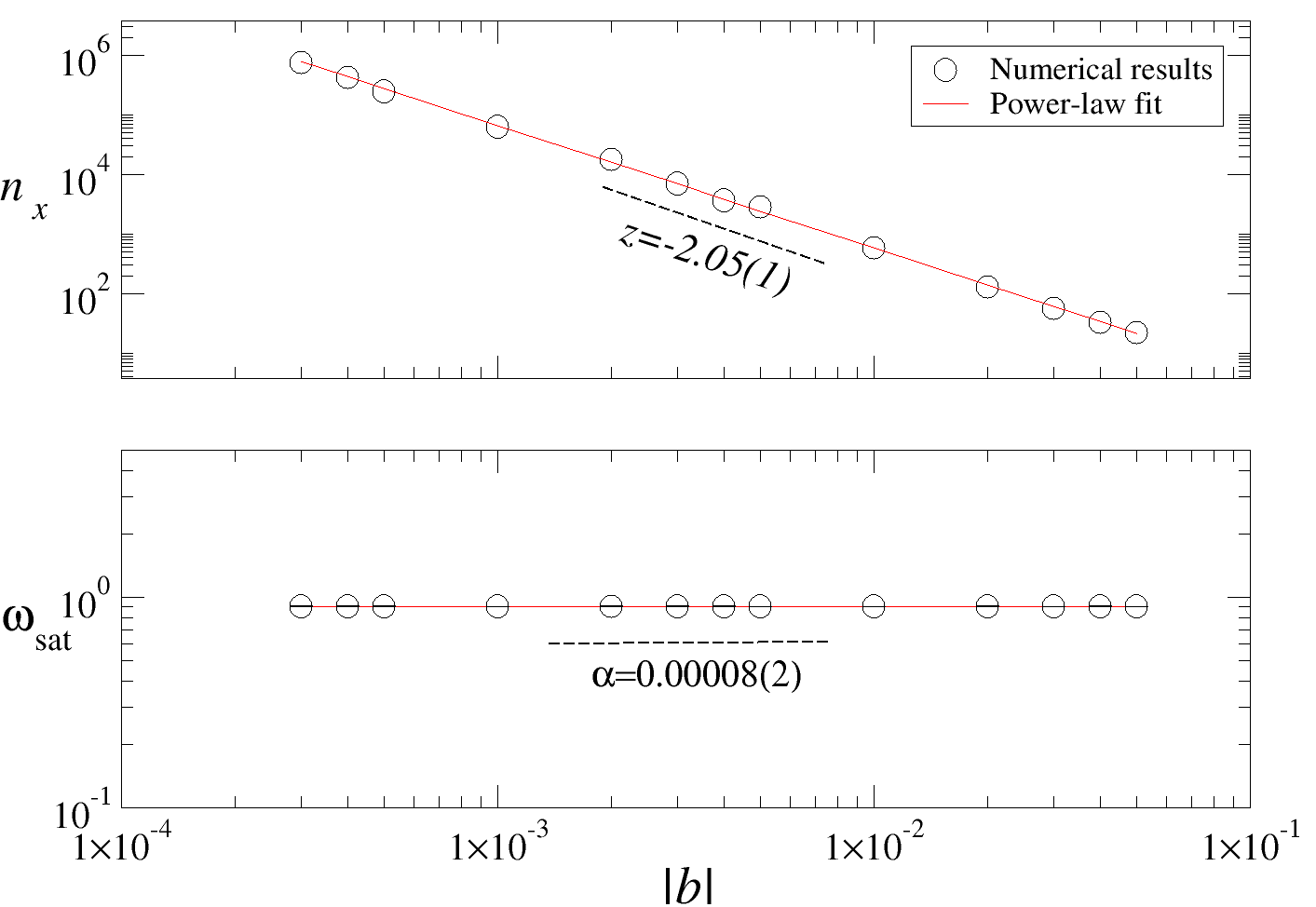}}
  \caption{Crossover iteration $n_x$ and stationary roughness $\omega_{\mathrm{sat}}$ as functions of $|b|$ in the dispersing regime. Power-law fits yield $z=-2.0(1)$ and $\alpha=0.0000(7)$, respectively.}
  \label{Fig8}
\end{figure}

The crossover iteration follows
\begin{equation}
n_x\propto |b|^z.
\end{equation}
Figure~\ref{Fig8} shows $n_x$ and $\omega_{\mathrm{sat}}$ as functions of $|b|$. The corresponding power-law fits yield
$z=-2.0(1)$ and $\alpha=0.0000(7)$, respectively. Moreover, the transformation $n\rightarrow n/|b|^z$ with $z=-2$ collapses the curves onto a single scaling function, as shown in Fig.~\ref{Fig7}(c).

The origin of these exponents can be understood directly from the map. In the dispersing regime, the angular increment is
\begin{equation}
\Delta\psi_n
=
\frac{8|b|}{a}(2\xi_{n+1}-1).
\label{EqDeltaPsi}
\end{equation}
For a strongly chaotic dynamics sampling $\xi$ approximately uniformly over $[0,1]$,
\begin{equation}
\langle 2\xi-1\rangle=0,
\qquad
\left\langle(2\xi-1)^2\right\rangle=\frac{1}{3}.
\end{equation}
Neglecting correlations between successive angular increments, the corresponding diffusion coefficient is therefore
\begin{equation}
D_\psi
\simeq
\frac{\langle(\Delta\psi)^2\rangle}{2}
=
\frac{32}{3a^2}b^2,
\label{EqDb}
\end{equation}
so that, independently of the prefactor,
\begin{equation}
D_\psi\propto b^2.
\end{equation}

This result establishes the connection with the confined stochastic diffusion discussed in Sec.~\ref{xsec2}. At the coarse-grained level, the angular dynamics is described by diffusion over the finite interval
$\psi\in[-\pi/2,\pi/2]$. Consequently, the stationary state is characterized by
\begin{equation}
\omega_{\mathrm{sat}}
=
\frac{\pi}{\sqrt{12}},
\end{equation}
giving $\alpha=0$, while normal diffusion implies $\beta=1/2$. Most importantly, the characteristic time required to explore a finite interval scales as
\begin{equation}
\tau_\psi
\sim
\frac{L_\psi^2}{D_\psi},
\end{equation}
where $L_\psi=\pi/2$ is independent of $b$. Since
$D_\psi\propto b^2$,
\begin{equation}
\tau_\psi\propto |b|^{-2}.
\end{equation}
Identifying the crossover iteration with this diffusive relaxation timescale therefore gives
\begin{equation}
n_x\sim\tau_\psi\propto |b|^{-2},
\end{equation}
and hence
\begin{equation}
z=-2,
\end{equation}
in quantitative agreement with the numerical result.

The deterministic billiard thus reproduces the complete exponent set of the confined stochastic model,
\begin{equation}
{
\alpha=0,\qquad
\beta=\frac{1}{2},\qquad
z=-2.
}
\end{equation}
The agreement does not rely on microscopic equivalence between the two systems. Rather, it results from the emergence of the same coarse-grained mechanism: normal diffusion with a perturbation-dependent diffusion coefficient $D\propto\epsilon^2$ inside a finite accessible domain. This common scaling structure provides the basis for identifying the stochastic random walk and the deterministic dispersing billiard with the same dynamical universality class.

\section{Characterizing the first-order phase transition}
\label{xsec4}

The results of the previous sections reveal a common dynamical scenario despite the fundamentally different microscopic nature of the two models. In both cases, the stationary diffusive observable changes discontinuously at the unperturbed limit, while the characteristic time required to reach the stationary regime diverges. We now formulate these observations within the language of phase transitions by focusing on two quantities: the macroscopic order parameter and the associated relaxation timescale.

In the theory of phase transitions, an order parameter is a macroscopic observable capable of distinguishing between different phases of a system \cite{landaunature,goldenfeld}. Its dependence on the control parameter provides a central criterion for identifying the nature of the transition. In a continuous transition, the order parameter approaches its critical value continuously, typically according to a power law. In a first-order transition, by contrast, it exhibits a finite discontinuity at the transition point \cite{goldenfeld}. In the present dynamical setting, an appropriate order parameter must quantify the extent to which the dynamics explores the available region of configuration or phase space.

We therefore identify the stationary values $x_{\mathrm{sat}}$ for the confined random walk and $\omega_{\mathrm{sat}}$ for the dispersing billiard as the corresponding macroscopic order parameters. For the stochastic model, when $\varepsilon=0$ the particle remains permanently at its initial position and therefore $x_{\mathrm{sat}}=0$. For every finite $\varepsilon\neq0$, however, the walker eventually explores the entire interval $[-L,L]$, yielding
\begin{equation}
x_{\mathrm{sat}}=\frac{L}{\sqrt{3}}.
\end{equation}
Accordingly,
\begin{equation}
x_{\mathrm{sat}}(\varepsilon)=
\begin{cases}
0, & \varepsilon=0,\\[2mm]
L/\sqrt{3}, & \varepsilon\neq0,
\end{cases}
\end{equation}
and therefore
\begin{equation}
\lim_{\varepsilon\rightarrow0^+}x_{\mathrm{sat}}(\varepsilon)
=
\frac{L}{\sqrt{3}}
\neq
x_{\mathrm{sat}}(0).
\label{EqJumpRW}
\end{equation}
The order parameter thus exhibits a finite jump at the transition.

An analogous discontinuity occurs in the dispersing billiard. At $b=0$, the rectangular geometry is integrable and the angle $\psi$ is conserved, so that the angular roughness vanishes. For any finite dispersing deformation, $b<0$, the dynamics becomes chaotic and eventually explores the accessible angular interval, yielding
\begin{equation}
\omega_{\mathrm{sat}}
=
\frac{\pi}{\sqrt{12}}.
\end{equation}
Consequently,
\begin{equation}
\lim_{b\rightarrow0^-}\omega_{\mathrm{sat}}(b)
=
\frac{\pi}{\sqrt{12}}
\neq
\omega_{\mathrm{sat}}(0),
\label{EqJumpBilliard}
\end{equation}
again revealing a finite discontinuity of the stationary observable at the critical point.

\begin{figure*}[t!]
  \centering
    \centerline{\includegraphics[width=0.7\textwidth]{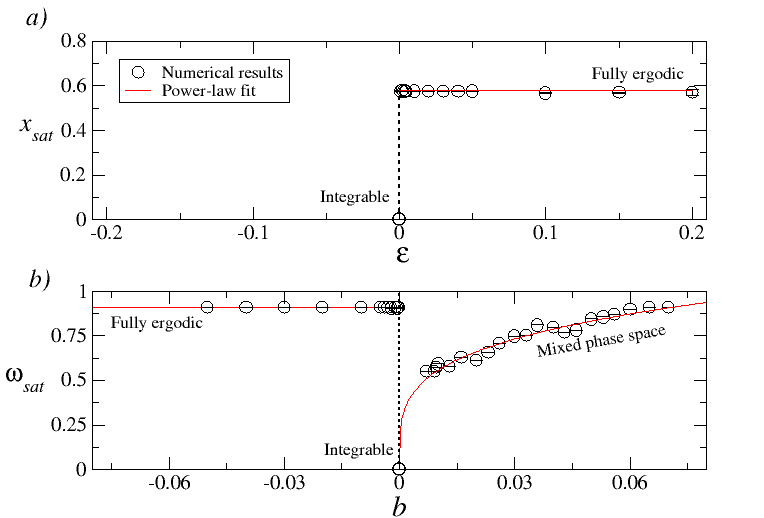}}
  \caption{Dependence of the macroscopic order parameters on the corresponding control parameters. (a) Stationary value $x_{\mathrm{sat}}$ as a function of the step amplitude $\varepsilon$ for the confined random walk. The order parameter jumps from $x_{\mathrm{sat}}=0$ at $\varepsilon=0$ to $L/\sqrt{3}$ for any finite $\varepsilon$. (b) Stationary roughness $\omega_{\mathrm{sat}}$ as a function of the geometric deformation $b$ for the stadium-like billiard. On the dispersing side ($b<0$), the order parameter exhibits a discontinuous jump at $b=0$, whereas on the focusing side ($b>0$) it approaches the integrable limit continuously, characterizing the second-order transition discussed in Ref.~\cite{daFonseca2026}. Symbols denote numerical results with the corresponding error bars, while solid lines represent the respective scaling fits.}
  \label{Fig9}
\end{figure*}

These behaviors are summarized in Fig.~\ref{Fig9}. Panel (a) shows the discontinuous response of $x_{\mathrm{sat}}$ to the stochastic step amplitude $\varepsilon$. Panel (b) reveals an even richer situation for the stadium-like billiard. Approaching $b=0$ from the dispersing side, $\omega_{\mathrm{sat}}$ remains finite and jumps discontinuously to zero at the integrable limit. Approaching the same point from the focusing side, however, $\omega_{\mathrm{sat}}$ vanishes continuously according to the scaling behavior discussed in Ref.~\cite{daFonseca2026}. The same integrable geometry therefore separates two qualitatively distinct routes away from integrability: a continuous transition toward a mixed phase space and a discontinuous transition toward a globally chaotic regime.

The discontinuity of the stationary observable is accompanied by a second important feature: a divergent dynamical timescale. Although the stationary value is independent of the perturbation amplitude on the discontinuous side of the transition, the time required to reach this state increases without bound as the critical point is approached. For both models, the diffusion coefficient vanishes quadratically with the perturbation,
\begin{equation}
D\propto\varepsilon^2
\qquad\text{or}\qquad
D_\psi\propto b^2,
\end{equation}
while the size of the accessible domain remains finite. Since diffusion across a finite domain requires a characteristic time of order
\begin{equation}
\tau\sim\frac{\ell^2}{D},
\end{equation}
where $\ell$ denotes the corresponding configuration- or phase-space scale, it follows that
\begin{equation}
\tau\propto\varepsilon^{-2}
\end{equation}
for the random walk and
\begin{equation}
\tau\propto |b|^{-2}
\end{equation}
for the dispersing billiard.

This analytical expectation is precisely reflected in the crossover iteration $n_x$ obtained numerically in Figs.~\ref{Fig2} and \ref{Fig8},
\begin{equation}
n_x\propto\varepsilon^{-2},
\qquad
n_x\propto |b|^{-2}.
\end{equation}
The crossover iteration can therefore be interpreted as a relaxation timescale controlling the approach to the stationary regime. Its divergence at the transition constitutes a dynamical manifestation of critical slowing down \cite{hohenberg}. Importantly, the divergence of the relaxation time does not require the stationary order parameter itself to vanish continuously. In the present case, a diverging timescale coexists with a finite jump of the asymptotic observable.

This combination distinguishes the discontinuous transition studied here from the continuous integrability-breaking transitions previously reported in nonlinear dynamical systems. On the discontinuous side, the stationary order parameter satisfies
\begin{equation}
\alpha=0,
\end{equation}
normal diffusion gives
\begin{equation}
\beta=\frac{1}{2},
\end{equation}
and the vanishing diffusion coefficient produces
\begin{equation}
z=-2.
\end{equation}
Thus, both the confined random walk and the deterministic dispersing billiard are characterized by the same exponent set
\begin{equation}
{
(\alpha,\beta,z)
=
\left(0,\frac{1}{2},-2\right).
}
\end{equation}

The physical origin of this agreement becomes transparent when the dynamics is viewed at a coarse-grained level. In both systems, transport is governed by normal diffusion within a finite accessible domain, while the corresponding diffusion coefficient vanishes quadratically as the transition is approached. The microscopic mechanisms generating the motion are entirely different -- stochastic independent increments in one case and deterministic chaotic scattering in the other -- yet they produce the same macroscopic scaling structure. The coincidence of the order-parameter behavior, relaxation scaling, and critical exponents therefore provides strong evidence that the two systems represent the same class of discontinuous dynamical transition.

The stadium-like billiard further reveals that continuous and discontinuous integrability-breaking transitions may emerge from opposite sides of the same integrable limit. For focusing boundaries, the chaotic component develops progressively and the order parameter vanishes continuously as $b\rightarrow0^+$. For dispersing boundaries, by contrast, an arbitrarily small deformation generates a globally accessible chaotic dynamics, producing a finite discontinuity as $b\rightarrow0^-$. This contrast establishes a unified dynamical setting in which first- and second-order transitions can be distinguished directly through the behavior of the stationary order parameter and the associated relaxation timescale.

\section{Summary and conclusions}
\label{xsec5}

We have investigated a discontinuous route from an unperturbed localized or integrable state to a diffusive chaotic regime using two dynamically distinct models: a confined one-dimensional stochastic random walk and a deterministic two-dimensional stadium-like billiard. In contrast with previously reported continuous integrability-breaking transitions, in which the stationary order parameter vanishes continuously as the control parameter approaches its critical value, the two systems considered here display a finite discontinuity of the corresponding stationary observable.

For the confined random walk, the stationary root-mean-squared displacement satisfies $x_{\mathrm{sat}}=0$ at $\varepsilon=0$, whereas for any finite $\varepsilon$ it approaches the geometry-controlled value $L/\sqrt{3}$. Likewise, in the dispersing billiard, $\omega_{\mathrm{sat}}=0$ at the integrable rectangular limit $b=0$, while an arbitrarily small dispersing deformation leads, at sufficiently long times, to the finite value $\pi/\sqrt{12}$. The discontinuity
\begin{equation}
\lim_{\lambda\rightarrow0}\mathcal{O}_{\mathrm{sat}}(\lambda)
\neq
\mathcal{O}_{\mathrm{sat}}(0),
\end{equation}
where $\lambda$ denotes the corresponding perturbation parameter, provides the central macroscopic signature of the first-order dynamical transition.

Despite this discontinuous stationary response, the approach to the asymptotic state becomes progressively slower near the transition. In both models, transport is governed by normal diffusion, characterized by $\beta=1/2$, while the effective diffusion coefficient vanishes quadratically with the perturbation amplitude. Because the size of the accessible domain remains finite, the associated relaxation timescale scales as $\tau\sim D^{-1}$, yielding
\begin{equation}
n_x\sim\tau\propto\lambda^{-2}.
\end{equation}
This explains the common crossover exponent $z=-2$ and provides a direct dynamical interpretation of the observed critical slowing down.

The two systems are therefore characterized by the same exponent set,
\begin{equation}
(\alpha,\beta,z)
=
\left(0,\frac{1}{2},-2\right),
\end{equation}
although their microscopic dynamics are fundamentally different. In the stochastic model, diffusion results from independent random increments, whereas in the billiard it emerges from deterministic chaotic scattering. Their agreement at the macroscopic level is thus not a consequence of microscopic equivalence, but of a common coarse-grained mechanism: normal diffusion within a finite accessible domain combined with a diffusion coefficient that vanishes quadratically at the transition.

The stadium-like billiard further provides a direct comparison between continuous and discontinuous routes away from the same integrable limit. On the focusing side, the progressive modification of the mixed phase space leads to a stationary observable that vanishes continuously as $b\rightarrow0$, as previously reported \cite{daFonseca2026}. On the dispersing side, by contrast, the stationary observable remains finite for arbitrarily small nonzero deformation and jumps discontinuously at $b=0$. The same integrable geometry therefore separates two distinct dynamical critical scenarios.

Taken together, these results support a statistical-mechanics description of integrability-breaking transitions based on macroscopic order parameters, scaling laws, and relaxation times. More importantly, they show that a discontinuous stationary response can coexist with a diverging dynamical timescale, providing a distinct signature of first-order dynamical criticality. The common scaling behavior observed in the stochastic and deterministic models further suggests that this mechanism is not restricted to a particular microscopic realization, but may define a broader class of discontinuous transitions in nonlinear dynamical systems.

\section*{Data Availability}
The data that support the findings of this study are available from the corresponding author upon reasonable request.
\begin{acknowledgments}
A.K.P.F. acknowledges CAPES (No.~88887.990665/2024-00), the Fulbright Program and the Fulbright Commission in Brazil -- Fulbright/CAPES Doctoral Dissertation Research Award (Process 2026--2027) for financial support. E.D.L. acknowledges support from CNPq (304398/2023-3) and FAPESP (2025/14544-0).
\end{acknowledgments}
\section*{Author Contributions}
\textbf{A.K.P.F.:} Conceptualization, Validation, Formal analysis, Investigation, Visualization, Writing – original draft. \textbf{M.A.P.:} Validation, Formal analysis, Investigation, Writing – review \& editing. \textbf{D.F.M.O.:} Methodology, Investigation, Writing – review \& editing, Supervision. \textbf{E.D.L.:} Methodology, Investigation, Writing – review \& editing, Supervision, Project administration.

\bibliography{PRE_draft}

@book{black,
  author    = {Black, Joseph},
  title     = {Lectures on the Elements of Chemistry, Delivered in the University of Edinburgh},
  editor    = {Robison, John},
  year      = {1803},
  publisher = {Printed by Mundell and Son, for Longman and Rees, London, and William Creech},
  address   = {Edinburgh}
}

@article{ehrenfest1933,
  author  = {Ehrenfest, Paul},
  title   = {Phasenumwandlungen im {\"u}blichen und erweiterten Sinn, klassifiziert nach den Singularit{\"a}ten des thermodynamischen Potentiales},
  journal = {Proceedings of the Royal Netherlands Academy of Arts and Sciences (Amsterdam)},
  volume  = {36},
  pages   = {153--157},
  year    = {1933}
}

@article{social,
title = {Social phase transitions},
journal = {Journal of Economic Behavior and Organization},
volume = {57},
number = {1},
pages = {71-87},
year = {2005},
issn = {0167-2681},
doi = {https://doi.org/10.1016/j.jebo.2003.11.013},
url = {https://www.sciencedirect.com/science/article/pii/S0167268104001490},
author = {Moshe Levy}
}

@article{landaunature,
  author  = {Landau, Lev D.},
  title   = {The Theory of Phase Transitions},
  journal = {Nature},
  volume  = {138},
  number  = {3498},
  pages   = {840--841},
  year    = {1936},
  month   = {Nov},
  doi     = {10.1038/138840a0}
}

@article{moreissame,
  author    = {Kadanoff, Leo P.},
  title     = {More is the Same; {P}hase {T}ransitions and {M}ean {F}ield {T}heories},
  journal   = {Journal of Statistical Physics},
  volume    = {137},
  number    = {5-6},
  pages     = {777--797},
  year      = {2009},
  month     = {Nov},
  doi       = {10.1007/s10955-009-9814-1},
  publisher = {Springer}
}

@article{bio1,
author = {Yongdae Shin  and Clifford P. Brangwynne },
title = {Liquid phase condensation in cell physiology and disease},
journal = {Science},
volume = {357},
number = {6357},
pages = {eaaf4382},
year = {2017},
doi = {10.1126/science.aaf4382},
URL = {https://www.science.org/doi/abs/10.1126/science.aaf4382}}

@article{bif,
title = {A mathematical framework for critical transitions: Bifurcations, fast–slow systems and stochastic dynamics},
journal = {Physica D: Nonlinear Phenomena},
volume = {240},
number = {12},
pages = {1020-1035},
year = {2011},
issn = {0167-2789},
doi = {https://doi.org/10.1016/j.physd.2011.02.012},
url = {https://www.sciencedirect.com/science/article/pii/S0167278911000443},
author = {Christian Kuehn}
}

@article{bio2,
  title = {Novel Type of Phase Transition in a System of Self-Driven Particles},
  author = {Vicsek, Tam\'as and Czir\'ok, Andr\'as and Ben-Jacob, Eshel and Cohen, Inon and Shochet, Ofer},
  journal = {Phys. Rev. Lett.},
  volume = {75},
  issue = {6},
  pages = {1226--1229},
  numpages = {0},
  year = {1995},
  month = {Aug},
  publisher = {American Physical Society},
  doi = {10.1103/PhysRevLett.75.1226},
  url = {https://link.aps.org/doi/10.1103/PhysRevLett.75.1226}
}

@article{MEISS,
title = {The destruction of tori in volume-preserving maps},
journal = {Communications in Nonlinear Science and Numerical Simulation},
volume = {17},
number = {5},
pages = {2108-2121},
year = {2012},
note = {Special Issue: Mathematical Structure of Fluids and Plasmas},
issn = {1007-5704},
doi = {https://doi.org/10.1016/j.cnsns.2011.04.014},
url = {https://www.sciencedirect.com/science/article/pii/S1007570411002085},
author = {J.D. Meiss}
}

@article{social2,
  title = {Disorder-induced phase transition in an opinion dynamics model: Results in two and three dimensions},
  author = {Mukherjee, Sudip and Chatterjee, Arnab},
  journal = {Phys. Rev. E},
  volume = {94},
  issue = {6},
  pages = {062317},
  numpages = {5},
  year = {2016},
  month = {Dec},
  publisher = {American Physical Society},
  doi = {10.1103/PhysRevE.94.062317},
  url = {https://link.aps.org/doi/10.1103/PhysRevE.94.062317}
}

@article{ehrenfest2,
  author  = {Sauer, Tilman},
  title   = {A look back at the {E}hrenfest classification: {T}ranslation and commentary of {E}hrenfest's 1933 paper introducing the notion of phase transitions of different order},
  journal = {The European Physical Journal Special Topics},
  volume  = {226},
  number  = {4},
  pages   = {539--549},
  year    = {2017},
  doi     = {10.1140/epjst/e2016-60344-y}
}

@book{gibbs,
  author    = {Gibbs, J. Willard},
  title     = {The Scientific Papers of J. Willard Gibbs, Volume 1: Thermodynamics},
  publisher = {Dover Publications Inc.},
  address   = {New York and London},
  year      = {1961}
}

@article{nontwist,
  title = {Ratchet current and scaling properties in a nontwist mapping},
  author = {Sales, Matheus Rolim and Borin, Daniel and de Souza, Leonardo Costa and Szezech, Jr., Jos{\'e} Danilo and Viana, Ricardo Luiz and Caldas, Iber{\^e} Luiz and Leonel, Edson Denis},
  journal = {Chaos, Solitons \& Fractals},
  volume = {189},
  pages = {115614},
  year = {2024},
  month = {dec},
  publisher = {Elsevier},
  doi = {10.1016/j.chaos.2024.115614}
}

@book{sethna2021statistical,
  title={Statistical mechanics: entropy, order parameters, and complexity},
  author={Sethna, James P},
  volume={14},
  year={2021},
  publisher={Oxford University Press, USA}
}

@article{Sinai70,
doi = {10.1070/RM1970v025n02ABEH003794},
year = {1970},
month = {apr},
publisher = {},
volume = {25},
number = {2},
pages = {137},
author = {Yakov G Sinai},
title = {Dynamical systems with elastic reflections},
journal = {Russian Mathematical Surveys}
}

@article{11nova,
  title = {Fermi-Ulam accelerator model under scaling analysis.},
  author = {Leonel, E D and McClintock, P V and Da Silva, J K},
  journal = {Physics Review Letters},
  volume = {93},
  number = {1},
  pages = {029902},
  year = {2004},
  doi = {10.1103/PhysRevLett.93.014101},
}

@article{Bunimovich1979,
  title = {On the ergodic properties of nowhere dispersing billiards},
  author = {Bunimovich, Leonid A.},
  journal = {Communications in Mathematical Physics},
  volume = {65},
  number = {3},
  pages = {295--312},
  year = {1979},
  publisher = {Springer},
  doi = {10.1007/BF01197884}
}

@article{khanna1991magnetic,
  title={Magnetic behavior of clusters of ferromagnetic transition metals},
  author={Khanna, SN and Linderoth, S{\o}ren},
  journal={Physical Review Letters},
  volume={67},
  number={6},
  pages={742},
  year={1991},
  publisher={APS}
}

@article{grinstein1976ferromagnetic,
  title={Ferromagnetic phase transitions in random fields: the breakdown of scaling laws},
  author={Grinstein, G},
  journal={Physical Review Letters},
  volume={37},
  number={14},
  pages={944},
  year={1976},
  publisher={APS}
}

@article{bianchi2002first,
  title={First-Order Superconducting Phase Transition in C e C o I n 5},
  author={Bianchi, A and Movshovich, Roman and Oeschler, N and Gegenwart, Philipp and Steglich, F and Thompson, Joe David and Pagliuso, PG and Sarrao, John Louis},
  journal={Physical Review Letters},
  volume={89},
  number={13},
  pages={137002},
  year={2002},
  publisher={APS}
}

@article{vojta2000quantum,
  title={Quantum phase transitions in d-wave superconductors},
  author={Vojta, Matthias and Zhang, Ying and Sachdev, Subir},
  journal={Physical Review Letters},
  volume={85},
  number={23},
  pages={4940},
  year={2000},
  publisher={APS}
}

@article{leonel2020characterization,
  title={Characterization of a continuous phase transition in a chaotic system},
  author={Leonel, Edson D and Yoshida, Makoto and de Oliveira, Juliano Antonio},
  journal={Europhysics Letters},
  volume={131},
  number={2},
  pages={20002},
  year={2020},
  publisher={IOP Publishing}
}

@article{leonel2015dynamical,
  title={A dynamical phase transition for a family of Hamiltonian mappings: A phenomenological investigation to obtain the critical exponents},
  author={Leonel, Edson D and Penalva, Julia and Teixeira, Riv{\^a}nia MN and Costa Filho, Raimundo N and Silva, M{\'a}rio R and De Oliveira, Juliano A},
  journal={Physics Letters A},
  volume={379},
  number={32-33},
  pages={1808--1815},
  year={2015},
  publisher={Elsevier}
}

@article{loskutov2002,
  title={Particle Dynamics in Time-Dependent Stadium-Like Billiards},
  author={Loskutov, A. and Ryabov, A.},
  journal={Journal of Statistical Physics},
  volume={108},
  pages={995--1014},
  year={2002},
  publisher={Springer},
  doi={10.1023/A:1019735313330}
}

@book{pathria2011statistical,
  title={Statistical Mechanics},
  author={Pathria, RK and Beale, Paul D},
  year={2011},
  publisher={Butterworth-Heinemann}
}

@article{oliveira2013some,
  title={Some dynamical properties of a classical dissipative bouncing ball model with two nonlinearities},
  author={Oliveira, Diego FM and Leonel, Edson D},
  journal={Physica A: Statistical Mechanics and its Applications},
  volume={392},
  number={8},
  pages={1762--1769},
  year={2013},
  publisher={Elsevier}
}

@article{kenji,
  title={A short review of phase transition in a chaotic system},
  author={Miranda, Lucas K. A. and Kuwana, Célica M. and Huggler, Yoná H. and da Fonseca, Anne K. P. and Yoshida, Makoto and Oliveira, Juliano A. and Leonel, Edson D.},
  journal={The European Physical Journal Special Topics},
  volume={231},
  pages={167–177},
  year={2022},
  publisher={Springer}
}

@article{kuehn2011mathematical,
  author  = {Kuehn, Christian},
  title   = {A mathematical framework for critical transitions: Bifurcations, fast-slow systems and stochastic dynamics},
  journal = {Physica D: Nonlinear Phenomena},
  volume  = {240},
  number  = {12},
  pages   = {1020--1035},
  year    = {2011},
  doi     = {10.1016/j.physd.2011.02.012}
}

@article{randomwalk,
  author   = {Pearson, Karl},
  title    = {The Problem of the Random Walk},
  journal  = {Nature},
  year     = {1905},
  volume   = {72},
  number   = {1865},
  pages    = {294--294},
  month    = {07},
  doi      = {10.1038/072294b0},
  url      = {https://doi.org/10.1038/072294b0}}

@article{scheffer2009early,
  author  = {Scheffer, Marten and Bascompte, Jordi and Brock, William A. and Brovkin, Victor and Carpenter, Stephen R. and Dakos, Vasilis and Held, Hermann and van Nes, Egbert H. and Rietkerk, Max and Sugihara, George},
  title   = {Early-warning signals for critical transitions},
  journal = {Nature},
  volume  = {461},
  number  = {7260},
  pages   = {53--59},
  year    = {2009},
  doi     = {10.1038/nature08227}
}

@article{vannes2007slow,
  author  = {van Nes, Egbert H. and Scheffer, Marten},
  title   = {Slow Recovery from Perturbations as a Generic Indicator of a Nearby Catastrophic Shift},
  journal = {The American Naturalist},
  volume  = {169},
  number  = {6},
  pages   = {738--747},
  year    = {2007},
  doi     = {10.1086/516845}
}

@article{livorati2011family,
  title={A family of stadium-like billiards with parabolic boundaries under scaling analysis},
  author={Livorati, Andr{\'e} LP and Loskutov, Alexander and Leonel, Edson D},
  journal={Journal of Physics A: Mathematical and Theoretical},
  volume={44},
  number={17},
  pages={175102},
  year={2011},
  publisher={IOP Publishing}
}

@book{chernov2006chaotic,
  title={Chaotic billiards},
  author={Chernov, Nikolai and Markarian, Roberto},
  number={127},
  year={2006},
  publisher={American Mathematical Soc.}
}

@article{daFonseca2025PRE,
  title = {Transition from bounded to unbounded energy in a time-dependent billiard},
  author = {da Fonseca, Anne K{\'e}tri P. and Silveira, Felipe Augusto O. and Kuwana, C{\'e}lia M. and Oliveira, Diego F. M. and Leonel, Edson D.},
  journal = {Phys. Rev. E},
  volume = {111},
  issue = {5},
  pages = {054215},
  year = {2025},
  month = {May},
  publisher = {American Physical Society},
  doi = {10.1103/PhysRevE.111.054215},
  url = {https://link.aps.org/doi/10.1103/PhysRevE.111.054215}
}

@article{denisnovo,
  title = {Describing a universal critical behavior in a transition from order to chaos},
  author = {Leonel, Edson D. and de Almeida, Mayla A. M. and Tarigo, Juan Pedro and Marti, Arturo C. and Oliveira, Diego F. M.},
  journal = {Phys. Rev. E},
  volume = {113},
  issue = {5},
  pages = {054220},
  numpages = {14},
  year = {2026},
  month = {May},
  publisher = {American Physical Society},
  doi = {10.1103/7231-j7zv}
}

@article{hohenberg,
  title={Theory of dynamic critical phenomena},
  author={Hohenberg, Pierre C and Halperin, Bertrand I},
  journal={Reviews of Modern Physics},
  volume={49},
  number={3},
  pages={435},
  year={1977},
  publisher={APS}
}

@book{goldenfeld,
  title={Lectures on Phase Transitions and the Renormalization Group (Frontiers in Physics)},
  author={Goldenfeld, Nigel},
  year={1992},
  publisher={Addison-Wesley},
  address={Reading, Massachusetts}
}

@misc{daFonseca2026,
      title={Integrability-breaking phase transitions in stadium-like billiards}, 
      author={da Fonseca, Anne K{\'e}tri P. and Leonel, Edson D.},
      year={2026},
      eprint={arXiv:2607.16482[nlin.CD]},
      archivePrefix={arXiv},
      primaryClass={nlin.CD}
}

@article{lopac2002chaotic,
  title={Chaotic dynamics and orbit stability in the parabolic oval billiard},
  author={Lopac, Vjera and Mrkonji{\'c}, Ivana and Radi{\'c}, Danko},
  journal={Physical Review E},
  volume={66},
  number={3},
  pages={036202},
  year={2002},
  publisher={APS}
}

\end{document}